"Keep up your bright swords, for the dew will rust them" – Othello

# Measurement of corrosion content of archaeological lead artifacts by their Meissner response in the superconducting state; a new dating method

S. Reich[*], G. Leitus[*] and S. Shalev[≠][•]

[*]Department of Materials and Interfaces and

[•]The Helen and Martin Kimmel Center for Archaeological Science

Weizmann Institute of Science

and

[≠]University of Haifa

**Meissner fraction in the superconducting state of lead archaeological artifacts is used to evaluate the mass of the uncorroded metal in the sample. Knowing the total mass of the sample the mass of all corrosion products is established. It is shown that this mass correlates with the archaeological age of the lead artifacts over a time span of ~2500 years. Well-dated untreated lead samples from Tel-Dor, the Persian period, Caesarea, the Byzantine and the Crusader periods as well as contemporary data were used to establish the dating correlation. This new chemical dating method is apparently applicable to lead artifacts buried in soils with the pH>6.5. In such soils the corrosion process is very slow and the corrosion products, mainly $PbO$ and $PbCO_3$, accumulate over hundreds of years. The method presented is in principle non-destructive.**

**Introduction**

Dating of metal artifacts is in most cases an unsolved problem. We propose a new chemical dating method for lead which is found quite frequently in archaeological excavations. In our method the corrosion extent on the surface of lead, used in antiquity for weights, pipes, claddings etc., is correlated with the burial time of the archaeological artifact. Conceptually our method is similar to the obsidian dating developed in 1960 by Irving Friedman & Robert L. Smith, see also Aitken, (1990) for review on chemical methods.

Lead metal at room temperature exhibits diamagnetic susceptibility of the same order of magnitude as its salts and oxide. When however cooled below 7.2K it enters into the superconducting state in which it exhibits the ideal diamagnetic susceptibility. The value of this susceptibility is $10^4$-$10^5$ higher than that of its oxide and salts which are not superconductors (s.c.). The value of the volume susceptibility ($\chi_v$) of lead in the Meissner state (see Meisner & Ochsenfeld, 1933), i.e. a state in which there is a total expulsion of the magnetic field from the volume of the lead sample, is $-1/4\pi$=-0.0796. This value translates to a molar susceptibility ($\chi_M$) of: $\chi_M$=-1.453 cgs. The molar susceptibility of lead oxide and lead salts is of the order of $-10^{-5}$ to $-10^{-4}$ cgs at room temperature (Handbook of Chemistry and Physics) and down to cryogenic temperatures. Thus the contribution to the diamagnetic signal of these compounds at cryogenic temperatures is negligible in comparison to the diamagnetic signal of lead. The volume magnetic susceptibility measured in the s.c. state of a corroded lead sample is:

$$-1/4\pi = \frac{M}{H_0} \times \frac{\rho}{m_{Pb}}(1-\eta) \qquad (1)$$

where M is the magnetic moment in emu; $\rho$=11.34 g/cm$^3$ the density of lead, $m_{Pb}$ is the mass of the lead metal in the sample and $\eta$ is the demagnetizing factor. $\eta$ is shape dependent and its measurement is presented later. $H_0$ is the value of the external magnetic field. Substituting the measured experimental values of M, $H_0$ and η into Eq. (1) we solve for the mass of the lead metal $m_{Pb}$. We use thus the magnetic measurement of the corroded sample to determine the mass of the uncorroded lead. This procedure yields the total mass of lead in the sample and does not depend on the spatial distribution of the corrosion materials in the sample their connectivity and shape.

Knowing the total mass of the sample, m, the mass $m_c$ of the corrosion products is determined:

$$m_c = m - m_{Pb} \qquad (2)$$

Experimentally we find that this procedure has the accuracy of ~1% in samples of ~100 mg.

By measuring the transition temperature to the s.c. state we can distinguish between relatively pure lead metal phase, $T_c$=7.2K, and homogeneous lead alloy phases which exhibit different values of $T_c$. Inclusions of a diamagnetic non-superconducting phase in the lead phase will not affect the $T_c$ value.

Lead's corrosion resistance accounts for its successful use in many environments when exposed to the atmosphere, soil fresh water and sea water (see Blaskett & Boxall, 1990). Its good resistance in contact with many corrodents is due to the formation of relatively insoluble films of corrosion products deposited on the metal in the initial stages of exposure, which then protect it from further attack. The resistance to corrosion of lead in soils is usually good and it varies with the acidity of the soil. In a controlled corrosion experiment of lead pipes in a number of soils for periods up to 10 years (see Gilbert, 1946) it has been found that for soils with pH above 6.5 the rate of corrosion was between $4 \times 10^{-5} cm/yr. - 6 \times 10^{-5} cm/yr$. In acidic soils pH 4-5 the rate was about fifty times higher. Thus we may assume that lead buried in sites with similar low value of acidity will have a comparable rate of corrosion over very long time periods.

The corrosion products of lead in soil are mainly lead oxide, PbO, and lead carbonate, $PbCO_3$. The solubility of these compounds in water at room temperature is very low; $2.3 \times 10^{-3}$ g/100ml and $1.1 \times 10^{-4}$ g/100ml respectively, thus the accumulation of these products on the surface of archaeological artifacts in soil over hundreds of years could be expected.

**Experimental and Results**

In our study the sampling of lead artifacts is done by punching out disks in a corroded untreated lead sheet or ribbon. These disks are 0.5-2.5mm in thickness and are 4.0mm in diameter. The punching procedure is performed with a hard steel shearing cylindrical knife.

Before the disk is cut, the corroded lead sheet is sandwiched between two paraffin films, which keep the brittle corrosion encapsulated on the sample. After the punching procedure the mass of the disk is measured. A SEM image of the surface of a sample from the Persian period is presented in Fig. 1. Note the $PbCO_3$ plate like crystallites on the surface. XRD spectra taken from the surface of the artifact from the Persian period show the following composition: 80 vol.% of cerussite, $PbCO_3$, 5 vol.% of litharge, PbO, the rest are inclusions in the corrosion layer of $SiO_2$ and some other minerals. After scraping off the corrosion layer from the lead artifact and homogenenizing its content in a mortar XRD spectrum was measured again. This time we observe 20 vol.% of $PbCO_3$ and 75 vol.% of PbO. This result shows that the surface of the corroded lead artifact is preferentially covered with cerussite as the XRD technique probes at most 1μm deep into the artifact. The cerussite mineral, which is practically insoluble, provides a good barrier for the accumulation of litharge in the corrosion layer (Fig. 2).

Magnetization vs. field and vs. temperature for the disk samples is measured in a SQUID MPMS$_2$ magnetometer. The magnetization vs. field at 5K and vs. temperature in a field of $H_0$=10 Oe. for lead artifacts ~2500yr old, Persian period, excavated in Tel Dor are shown in Fig. 3. The critical field $H_c$ and the critical temperature, $T_c$, are the same as those measured in contemporary samples. To calculate the demagnetizing factor, $\eta$, the magnetization in small enough fields, $0 \leq H_0 \leq 30 Oe$, is measured in two configurations: disks parallel and perpendicular to $H_0$. Figure 4 presents data for fourteen disks cut out from seven artifacts.

It can be shown that for thin enough disks (see Poole, Farach & Creswick 1995):

$$m_{Pb} = -4\pi\rho\chi_{\exp\|} /(1+\chi_{\exp\|}/2\chi_{\exp\perp}) \qquad (3)$$

where $\chi_{\exp\|} = \dfrac{M_\|}{H_0}$; $\chi_{\exp\perp} = \dfrac{M_\perp}{H_0}$

$$\eta_\| = \frac{1}{2}\frac{U}{1+\frac{1}{2}U}; \quad \eta_\perp = 1 - \frac{U}{1+\frac{1}{2}U}; \quad U \equiv \frac{\chi_{\exp\|}}{\chi_{\exp\perp}}$$

The $m_c/cm^2$ values for these disks as calculated from Eqs.(3) and (2) are shown in Table I. The archaeological site, the pH of the soil (see Thomas, 1996) at the site, the archaeological dating of the sampled objects (see Renfrow & Bahn, 1991) and an estimated corrosion time are presented for samples from the Persian, the Byzantine and the Crusader periods, contemporary samples are detailed in Table 1 as well.

In Fig. 5 we present the corrosion extent for the above samples as a function of the estimated corrosion time. The power law curve y(t)= $\dfrac{\overline{m}_c(t)}{cm^2} = kt^{\frac{1}{\alpha+1}}$, $\alpha$=0.07, is the best fit to the experimental points.

**Discussion**

A power law dependence of y on t is reasonable (see Chilton, 1968). In a very general way the act of corrosion is based on the local existence of an electrochemical cell, which comprises a junction of a metallic oxide to a metal surface on one hand and to oxygen and water on the other. This cell is characterized by certain potential difference, ε, that drives a current, I, through the cell:

$$\frac{dy}{dt} \sim I = \frac{\varepsilon}{R(y)}; \quad \text{assuming} \quad R(y) \sim y^\alpha \text{ we get} \quad y^\alpha dy = K_1 dt \rightarrow \frac{y^{\alpha+1}}{\alpha+1} = K_1 t + K_2$$

since at t=0 y=0 $\rightarrow K_2 = 0$ $\quad\quad y = Kt^{\frac{1}{\alpha+1}}; \quad K = (\alpha+1)K_1$.

The power law fit is close to a rectilinear line. Such a behavior is characteristic for a corrosion process in which a thin protective layer of oxide is formed on the surface of the metal, but when the stresses within the oxide become so great that when the layer reaches a certain thickness it breaks down and the metal surface is re-exposed to the environment. More oxide is then formed in contact with the metal, and, after a similar time interval, the thin film of corrosion is again disrupted, and then the process is repeated. The net result of such behavior is the production of an oxidation curve, which is virtually a straight line (see Chilton, 1968). This curve may serve for the approximate dating of lead archaeological artifacts buried in soils of relatively high pH, such as carbonate buffered soils, where the corrosion rate is comparable and relatively slow so that time averaging of external condition does occur.

As the relevant corrosion signal grows in time the relative dating accuracy should improve with the age of the sample. Thus the 2500 years span, dictated by the availability of samples, is not an upper limit of our method.

The new method presented here for the determination of lead metal content in a nonmagnetic artifact is in principle non-destructive. Thus for example lead inclusions in copper, silver and gold ancient coins can be studied, as well.

**Acknowledgements**

The authors wish to thank the archaeologists: A. Raban and Y. Arnon from Caesarea excavation, and E. Stern and A. Gilboa from Dor excavation, for providing suitable samples from well dated archaeological contexts.

**References**

Aitken M.J. (1990). *Science-based Dating in Archaeology*, Chapter 8, New York: Longman Inc.

Blaskett, D.R. & Boxall O. (1990). *Lead and its alloys*, England: Ellis Horwood Limited, pp. 128-140.

Chilton J.P. (1968). *Principles of Metallic Corrosion*,. London: The Royal Institute of Chemistry,. pp. 5-13.

Friedman I. & Smith R.L. (1960). *A new dating method using obsidian*, American Antiquity **25**, 476-493.

Gilbert P.T. (1946). *Corrosion of copper, lead, and lead-alloy specimens after burial in a number of soils for periods up to 10 years*, J. of Institute of Metals, pp. 139-174.

*Handbook of Chemistry and Physics*, E111, 69$^{th}$ edition.

Meissner W. & Ochsenfeld R. (1933). *Ein neuer effekt bei eintritt der Superleitfähigkeit*, Naturwiss. **21,** 787-788.

Poole C.P., Jr., Farach H.A. & Creswick R.J. (1995). *Superconductivity*, Chapters 9-12, New York: Academic Press.

Renfrew C. & Bahn P. (1991). *Archaeology: Theories, Methods and Practice*, Chapter 4, London: Thames and Hudson.

Thomas, G.W. (1996). In *Methods of Soil Analysis*, part 3, Chemical Methods, Chapter 16, Pub. Soil Science Society of America Inc.

**Figure Captions**

Fig. 1. SEM image of a corroded lead disk punched out from an archaeological artifact from the Persian period. $PbCO_3$ crystallites are present on the surface.

Fig. 2. XRD spectra; a: Spectrum from the corrosion on the surface of a disk, b: Spectrum from homogenized corrosion powder scraped off the surface of the disk. The sample is from the Persian period.

Fig. 3. a: Magnetization vs. field of a corroded lead disk from the Persian period measured at 5K. The disk is measured in two configurations; parallel and perpendicular to the external magnetic field. b: Magnetization vs. temperature for two configurations measured in an external magnetic field of 10 Oe in a zero field-cooled experiments.

Fig. 4. Magnetization of corroded disks from different periods. Disks are parallel to the external magnetic field and measured at 5K. The numbers refer to Table I. Note that the slopes of these straight lines is proportional to the mass of the lead metal in the samples and are inversely proportional to $(1-\eta_\parallel)$, where $\eta_\parallel$ is the demagnetizing factor for the parallel configuration.

Fig. 5. Average corrosion mass $\overline{m}_c$ per unit nominal surface area of the disks as function of time. The power law curve is the best fit to the experimental points.

Table I.

| Sample # | m, mg | $m_{Pb}$, mg | $m_c$, mg | $m_c/S$, mg/cm$^2$ | Shape of artifact | Archaeological site | Period | Approximate time of corrosion, years |
|---|---|---|---|---|---|---|---|---|
| 762 | 118.63 | 58.6 | 60.0 | 240.0 | Pb sheet corrosion on both sides | Tel Dor Soil's pH of the archaelogical stratum: 8.23 Surface site control: 8.17 | Persian | 2430±100 |
| 765 | 120.00 | 82.9 | 37.1 | 148.4 | | | - " - | - " - |
| 770 | 121.81 | 72.7 | 49.1 | 196.4 | | | - " - | - " - |
| 771 | 114.80 | 57.7 | 57.1 | 228.4 | | | - " - | - " - |
| 772 | 110.33 | 67.7 | 42.6 | 170.4 | | | - " - | - " - |
| 774 | 98.42 | 63.3 | 35.1 | 140.4 | | | - " - | - " - |
| 775 | 151.81 | 97.6 | 54.2 | 216.8 | | | - " - | - " - |
| | | | Average | 191.5 | | | | |
| | | | Standard error | 14.8 | | | | |
| 807 | 105.3 | 82.6 | 22.7 | 90.8 | Pb ribbon corrosion on both sides | Caesarea Soil's pH of the archaelogical stratum: 7.96 Surface site control: 8.51 | Byzantine | 1350±100 |
| 809 | 124.0 | 107.2 | 16.8 | 67.2 | | | - " - | - " - |
| 813 | 178.8 | 159.9 | 28.9 | 115.6 | | | - " - | - " - |
| | | | Average | 91.2 | | | | |
| | | | Standard error | 14.0 | | | | |
| 796 | 176.73 | 155.8 | 20.9 | 83.6 | Pb ribbon corrosion on both sides | Caesarea Soil's pH of the archaelogical stratum: 7.96 Surface site control: 8.51 | Crusader | 750±50 |
| 797 | 210.41 | 200.9 | 10.5 | 42.0 | | | - " - | - " - |
| 799 | 184.00 | 158.4 | 25.6 | 102.4 | | | - " - | - " - |
| 800 | 219.7 | 203.8 | 15.9 | 63.6 | | | - " - | - " - |
| | | | Average | 72.9 | | | | |
| | | | Standard error | 13.0 | | | | |
| | | | | 5.0 | Pb pipes | Data from Gilbert, 1946, soil's pH > 6.5 | Contem-porary | 10 |

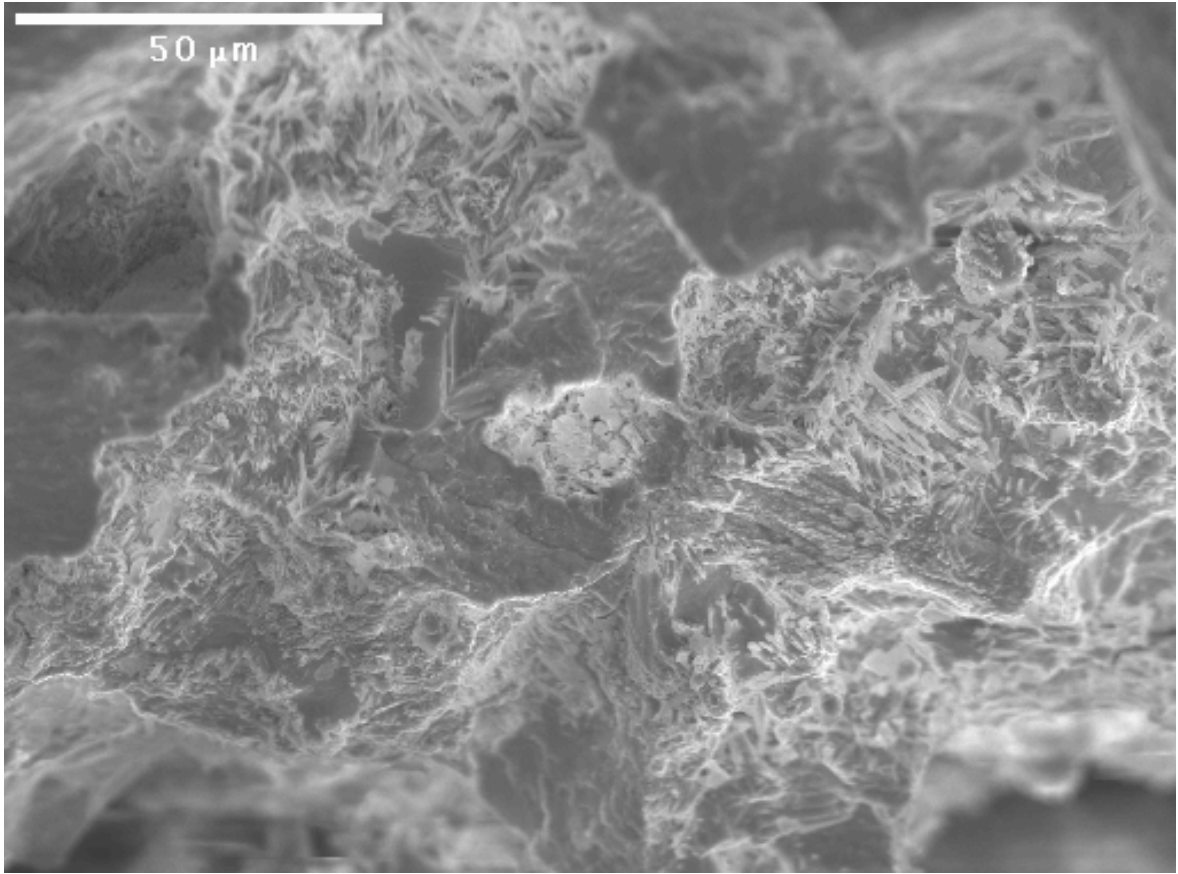

Figure 1.

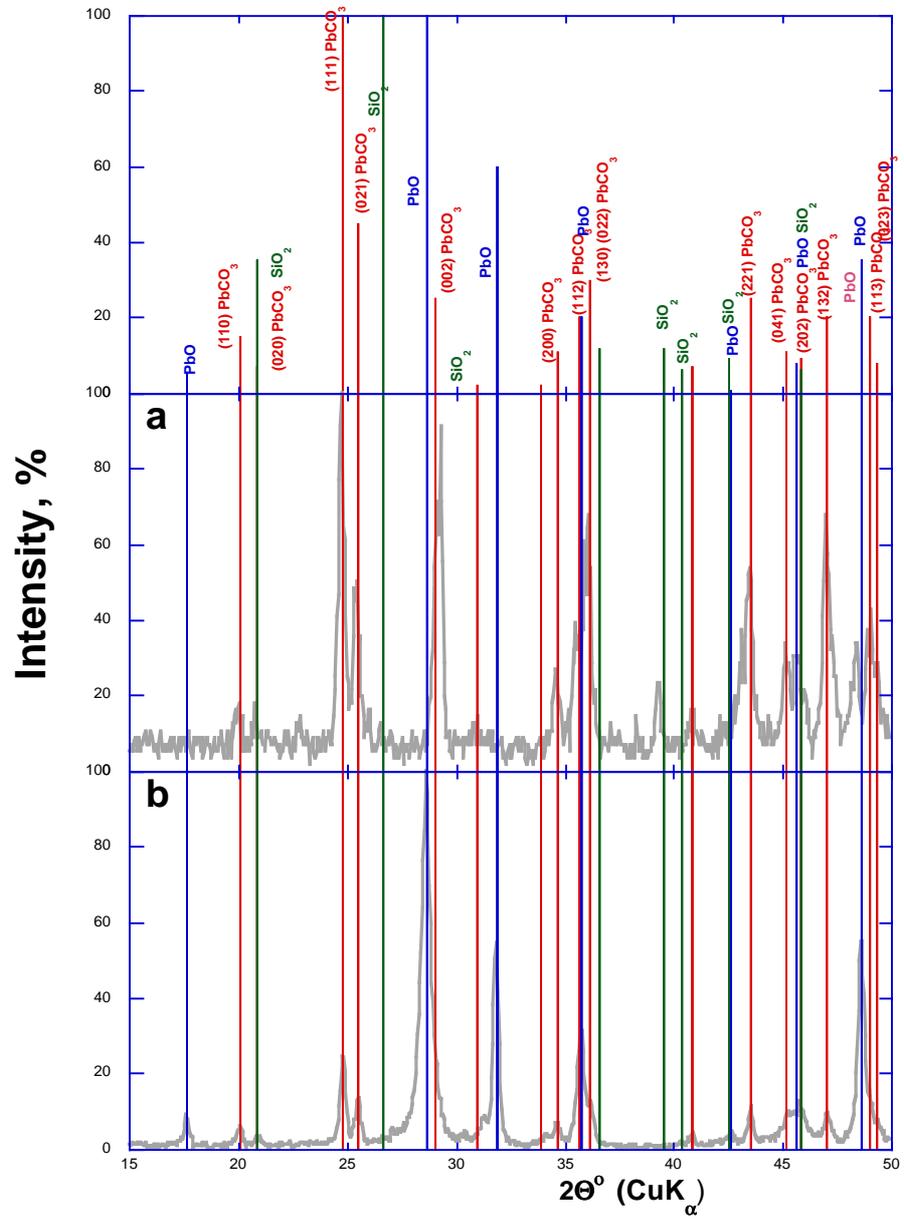

Figure 2.

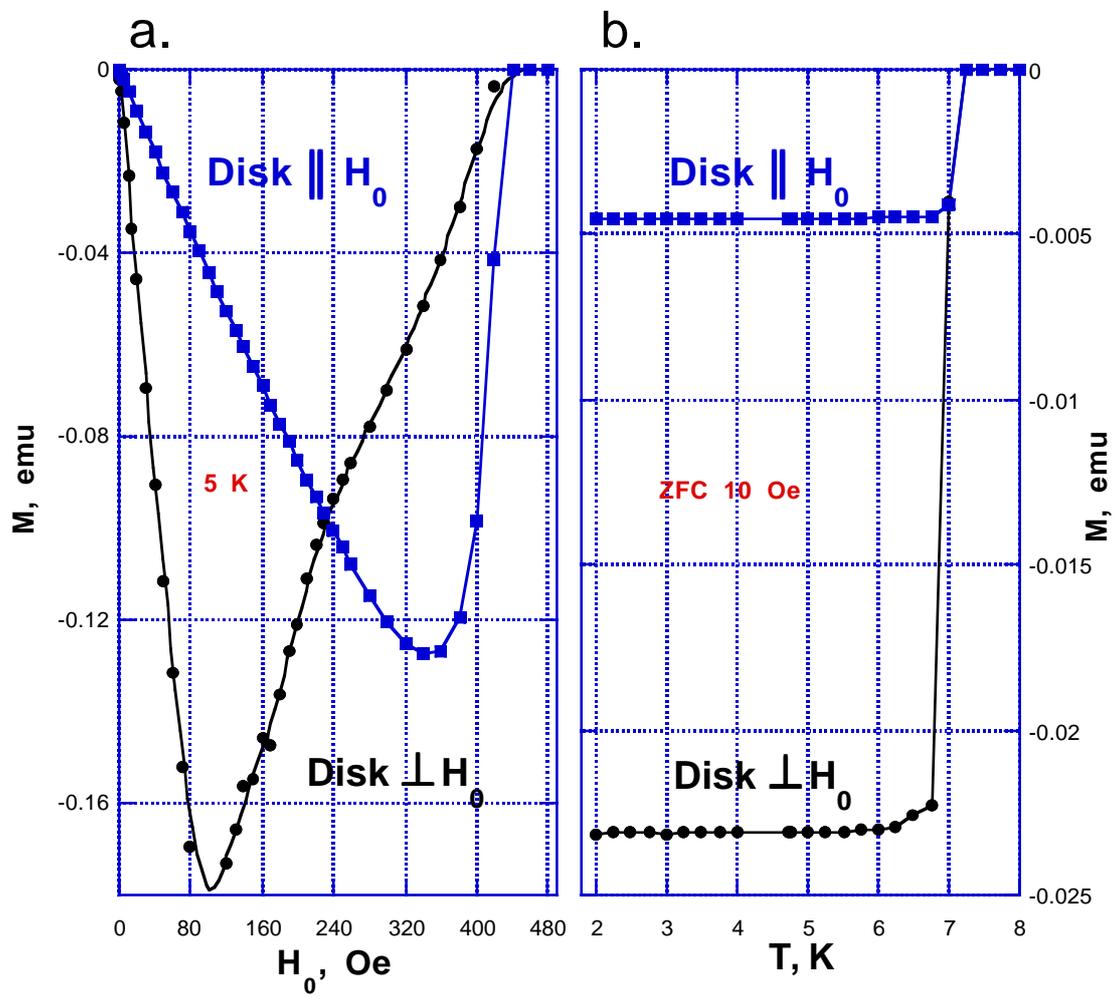

Figure 3.

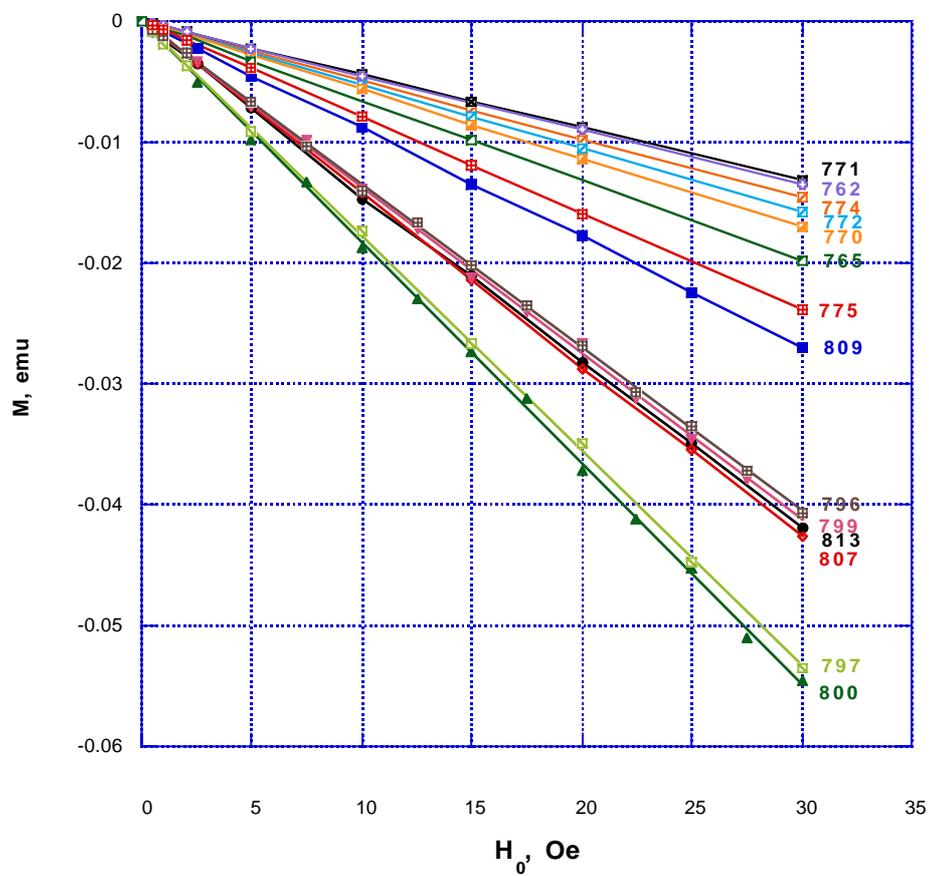

Figure 4.

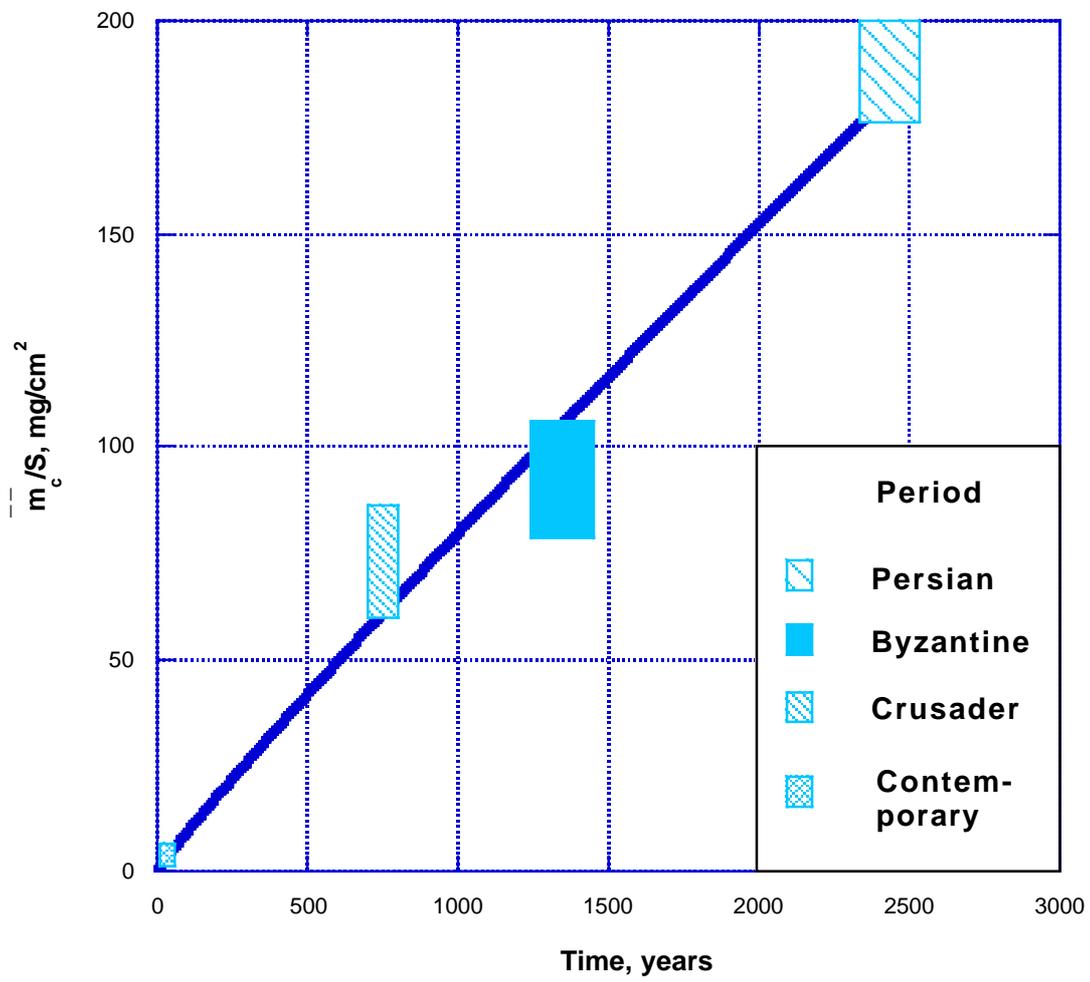

Figure 5.